\documentclass[english,showpacs,amsmath,twocolumn,prb]{revtex4}
\usepackage[T1]{fontenc}
\usepackage[latin9]{inputenc}
\usepackage{textcomp}
\usepackage{amstext}
\usepackage{graphicx}

\makeatletter

\providecommand{\tabularnewline}{\\}

\@ifundefined{textcolor}{}
{%
 \definecolor{BLACK}{gray}{0}
 \definecolor{WHITE}{gray}{1}
 \definecolor{RED}{rgb}{1,0,0}
 \definecolor{GREEN}{rgb}{0,1,0}
 \definecolor{BLUE}{rgb}{0,0,1}
 \definecolor{CYAN}{cmyk}{1,0,0,0}
 \definecolor{MAGENTA}{cmyk}{0,1,0,0}
 \definecolor{YELLOW}{cmyk}{0,0,1,0}
 }

\makeatother

\usepackage{babel}

\begin{document}

\title{Probing Aromaticity of Borozene Through Optical and Dielectric Response:
A Theoretical Study }

\author{Sridhar Sahu and Alok Shukla}

\address{Department of Physics, Indian Institute of Technology, Bombay, Powai,
Mumbai 400076, INDIA}

\email{sridhar@phy.iitb.ac.in, shukla@phy.iitb.ac.in}
\begin{abstract}
In this work, we report electronic structure calculations aimed at
computing the linear optical absorption spectrum, and static dipole
polariziblity of a newly proposed boron-based planar aromatic compound
borozene ($B_{12}H_{6}$). For the purpose, we use the semiempirical
INDO model Hamiltonian, accompanied by large-scale correlation calculations
using the multi-reference singles-doubles configuration-interaction
(MRSDCI) approach. We present detailed predictions about the energetics,
polarization properties, and the nature of many-particle states contributing
to various peaks in the linear absorption spectrum. Our results can
be used to characterize this material in future optical absorption
experiments. We also argue that one can deduce the aromaticity of
the cluster from the optical absorption and static polarizability
results.
\end{abstract}


\maketitle

\section{Introduction}

\label{sec:intro}

Next to carbon-based chemistry, boron chemistry is surely one of the
most active areas of research\cite{davidson-book}. Boron exhibits
a variety of polymorphisms which involve three-center two-electron
(3c-2e) bonds\cite{libscomb-book}, planar aromatic structures\cite{alexandrova},
nano-tubes\cite{ciuparu}, nano-ribbons\cite{xu} and fullerene-like
cages\cite{yakobson-fuller}. From the point of view of nano-technology
aromatic boron materials are of particular interest because delocalized
electron clouds in these materials will lead to large linear and non-linear
optical susceptibilities, just like corresponding carbon-based structures.
In order to achieve aromaticity, one needs to have a planar structure
giving rise to $\pi$-orbitals. Indeed, aromaticity in boron-based
clusters has been the subject of many recent investigations. Zhai
\emph{et al.\cite{wang-nat-mat} }in a recent joint theoretical and
experiment study probed the issue of planarity and aromaticity boron
clusters. In another theory-experiment study, Zhai \emph{et al.\cite{wang-angew-chem}
}reported the discovery of planar boron clusters B$_{8}$ and B$_{9}$
clusters with {}``molecular wheel'' like structures, and exhibiting
both $\sigma$ and $\pi$ aromaticity. Aihara and coworkers\cite{ishida-jacs}
presented a theoretical investigation of aromaticity in several planar
and quasi-planar boron clusters using the concept of {}``topological
resonance energy''. Johansson\cite{johannson-jphyschemc} demonstrated
theoretically the existence of strong magnetically-induced ring currents
B$_{20}$ and other toroidal clusters of boron, as an evidence of
underlying aromaticity. Rincon \emph{et al.}\cite{dalton-sigmaaromatic}
theoretically demonstrated the $\sigma$ aromaticity in several planar
boron clusters performing an analysis of the electron-localization
function (ELF). Similarly, Wu \emph{et al}.\cite{boro-carbon-aromatic}
using an \emph{ab initio} approach, investigated the aromaticity of
planar boron-carbon complex C$_{6}$B$_{\text{12}}$$^{\text{-2}}$.
Another possible approach to achieve planarity, and, therefore, aromaticity
in boron clusters is by hydrogenation of boron clusters. For example
Alexandrova \textit{et al}.\cite{alexandrova1} have demonstrated
by means of theoretical calculations, that quasi-planar $B_{7}^{-}$
cluster, upon hydrogenation, acquires a planar structure for $B_{7}H_{2}^{-}$.
In a recent theoretical work, Szwacki \emph{et al.}\cite{szwacki}
proposed a novel planar aromatic cluster which can be obtained by
hydrogenation of quasi-planar B$_{12}$ cluster. They argued that
the cluster is aromatic by performing a detailed analysis of its molecular
orbitals, and by examining magnetic properties such as nucleus-independent
chemical shift (NICS), and anisotropy of magnetic susceptibility (AMS).
Having thus demonstrated the similarities between the aromaticities
of this cluster and benzene (including the number of $\pi$-electrons),
they called it the boron analogue of benzene, and named it borozene\cite{szwacki}.
Encouraged by the work of Szwacki \emph{et al.}\cite{szwacki}, Forte
\textit{et al.}\cite{Forte} performed \emph{ab initio} calculations
to predict larger aromatic compounds which can be constructed using
borozene as formula unit. It is a well-known fact that the carbon-based
aromatic materials including molecules such as benzene, napthalene,
anthracene and longer $\pi$-conjugated polymers exhibit intense linear
and non-linear optical response\cite{barford-book}. Therefore, it
is of considerable interest to explore similar properties of aromatic
materials consisting of elements other than carbon. With this aim
in mind, here we present a systematic theoretical study, based upon
a large-scale configuration interaction (CI) methodology, of linear
optical response of borozene. We believe our results will be useful
in optical characterization experiments on this material, as also
in exploring its possible application in nano-optics. Furthermore,
we also perform calculations of static polarizability of this cluster,
with the aim of exploring the signatures of aromaticity in its dielectric
response. Aromaticiy is an intuitive concept which essentially implies
electron delocalization. Therefore, the consequences of this delocalization
of electrons will be manifest in various properties of materials,
including their magnetic as well as dielectric response. With the
electrons delocalized along the plane of the molecule for conjugated
systems as benzene, one would expect its {}``in-plane'' dielectric
response to be significantly more than its {}``perpendicular'' response.
Thus, this anisotropy in the static polarizability tensor $\alpha_{ij}$
can also be viewed as a signature of the aromaticity, in such systems.
Indeed, this anisotropy in $\alpha$ has been verified theoretically
by other authors in the past calculations on conjugated molecules,\cite{lazzeretti-alpha,perez-alpha,hinch-alpha}
and has been used to formulate an {}``aromaticity scale'' for such
systems.\cite{lazzeretti-alpha} 

Remainder of the paper is organized as follows. In section \ref{sec:intro}
we briefly describe the theoretical methodology employed for the present
calculations, and also discuss the geometry of the molecule. This
is followed by the presentation and discussion of our results in section
\ref{sec:results}. Finally, in section \ref{sec:conclusions} we
present our conclusions.

\section{Theoretical Methodology}

\label{sec:theory}

\begin{figure}
\includegraphics[angle=-1,width=6cm]{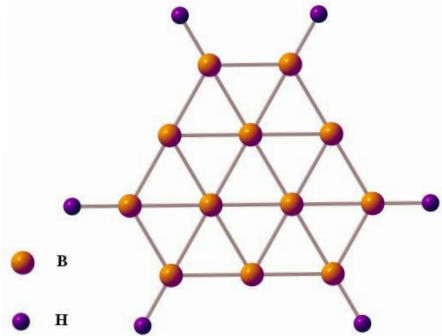} 

\caption{(Color online) Structure of S-borozene considered in this work, assumed
to lying in the $xy$-plane. Optimized bond lengths at INDO-HF level
are B-B = 1.65 \AA\  and B-H = 1.18 \AA. Yellow dots indicate boron
atoms and violet ones hydrogen atoms.}

\label{Fig-struct}
\end{figure}

The basic structure of borozene is presented in Fig. \ref{Fig-struct}
and it corresponds to point symmetry group $D_{3h}$\cite{szwacki}.
Before proceeding with our calculations, for the purpose of verification,
we decided to perform the geometry optimization on our own using the
B3LYP based hybrid DFT approach employing the 6-311++g(d) basis set
as implemented in GAUSSIAN03 program\cite{gaussian}. The optimized
structure thus obtained belonged to the $D_{3h}$ point group symmetry
with uniform value of B-H bond length $1.18$ \AA , and four distinct
values of B-B bond lengths $1.63$ \AA, $1.66$ \AA, $1.81$ \AA,
and $1.86$ \AA. We, however, are interested in exploring an even
more symmetric structure of borozene, so as to make the comparison
with benzene more transparent. Therefore, we separately optimized
a highly symmetric structure of borozene (S-borozene, henceforth)
with identical B-B bond lengths, in addition to the B-H ones. For
this purpose, as well as for the linear absorption spectrum calculations,
we adopted an INDO model Hamiltonian\cite{pople-indo} based approach
implemented recently in a computer program developed by us\cite{sahu-cpc}.
INDO model is an effective valence-electron approach, employing a
Slater-type minimal basis set, and some semi-empirical parameters\cite{pople-indo}.
The geometry optimization using our computer program, performed at
the Hartree-Fock (HF) level (INDO-HF, henceforth), yielded bond lengths
1.65 \AA\  for the B-B bond, and 1.18 \AA\  for the B-H bond.
These values agree perfectly with $1.18$ \AA\  for the B-H bond
length reported both by Szwacki \emph{et. al} \cite{szwacki} and
Forte \emph{et.} \emph{al} \cite{Forte}, and the average value of
B-B bond length $1.649$ \AA\  reported by Forte \emph{et.} \emph{al}
\cite{Forte}. The total energy of the geometry-optimized S-borozene
was higher than that of the borozene with distinct B-B bond lengths
by $\approx0.8$ eV at the INDO-HF level. 

Next, large-scale correlation calculations were performed on the ground
and the excited states of S-borozene, using the multi-reference singles-doubles
configuration-interaction (MRSDCI) approach as implemented in the
MELD package\cite{meld}, but employing the one- and two-electron
matrix elements of the INDO Hamiltonian supplied by our program\cite{sahu-cpc}.
Thus, these correlated calculations were entirely within the INDO
model, and will be called INDO-MRSDCI calculations, henceforth. During
these calculations, point symmetry group $C_{2v}$ was utilized ,
as against the $D_{3h}$ symmetry group because the MELD\cite{meld}
package is restricted to $D_{2h}$ and its sub-groups. Therefore,
we classify the many-electron states of the system in terms of the
irreducible representations (irreps) of $C_{2v}$ group. The optical
absorption spectra was calculated under the electric-dipole approximation
employing the Lorenzian line shape. For both the ground and the excited
states, these calculations were performed in an iterative manner,
until the optical absorption spectra computed from them converged.
We have extensively used this approach in our earlier calculations
on conjugated polymers\cite{mrsd-calc} as well as on B$_{\text{12}}$
icosahedral and quasi-planar clusters reported recently\cite{sahu-b12}.

\section{Results and Discussion}

\label{sec:results}

\begin{figure}
\includegraphics[width=4cm]{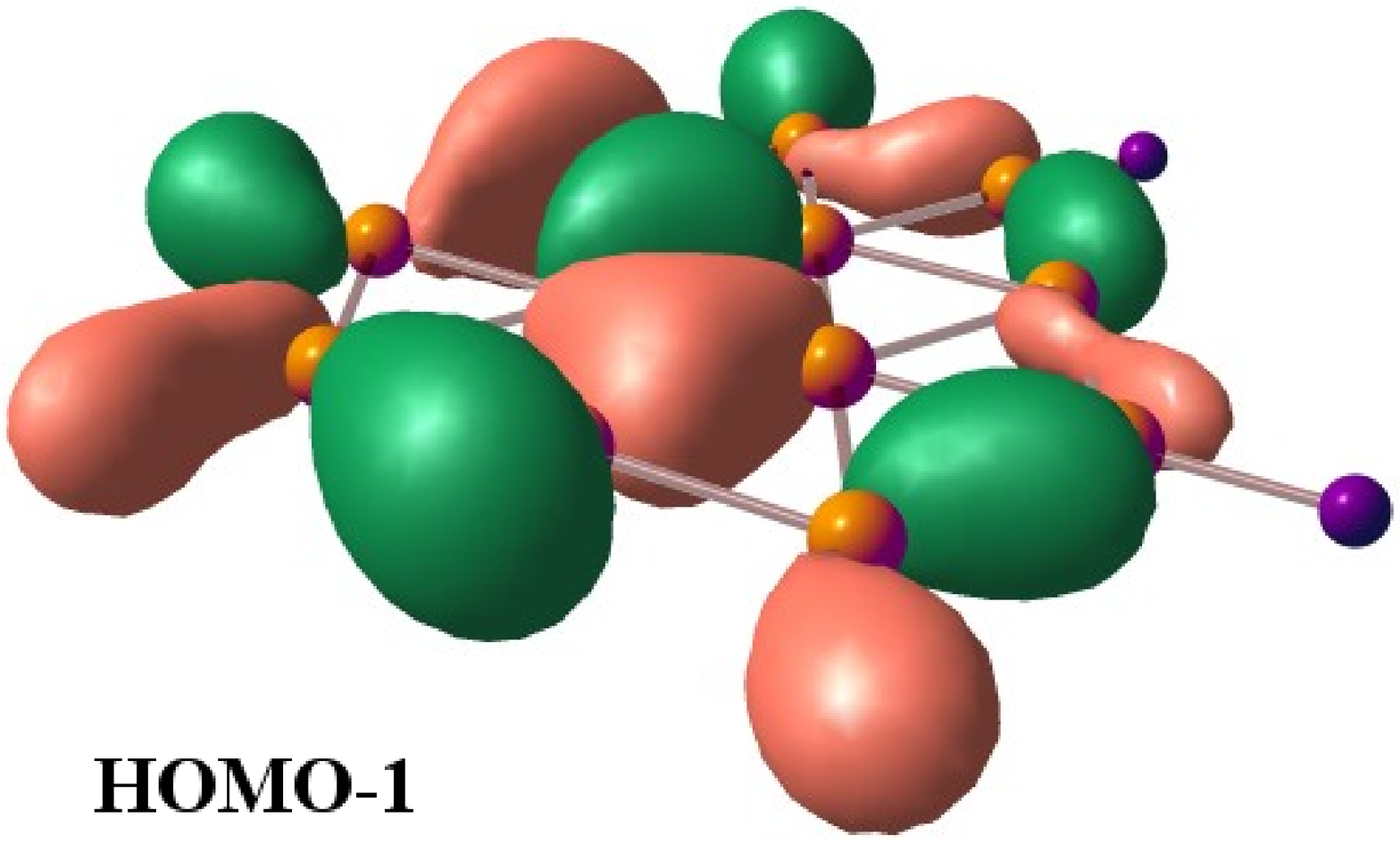}\includegraphics[width=4cm]{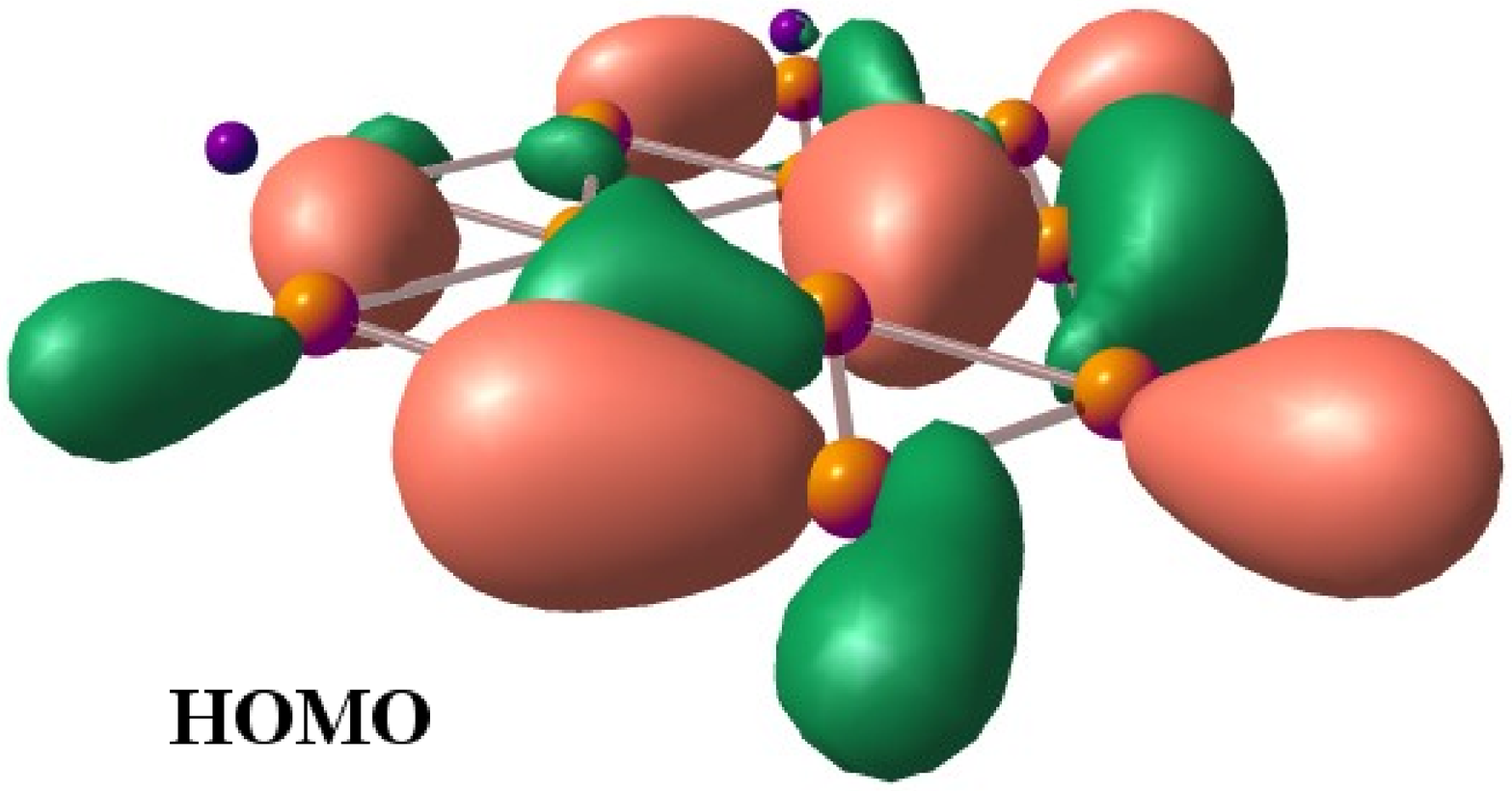}

\bigskip{}

\includegraphics[width=4cm]{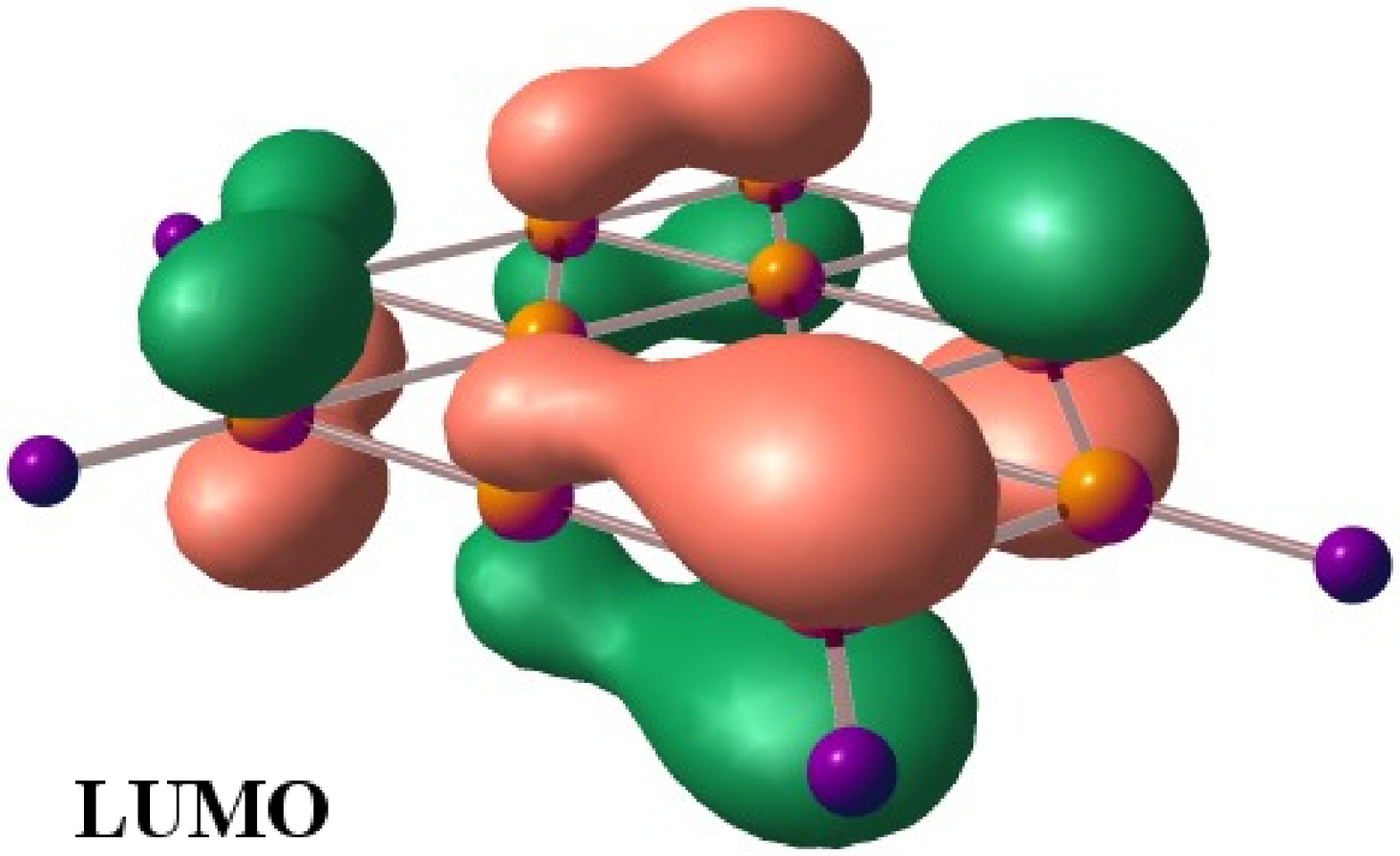}\includegraphics[width=4cm]{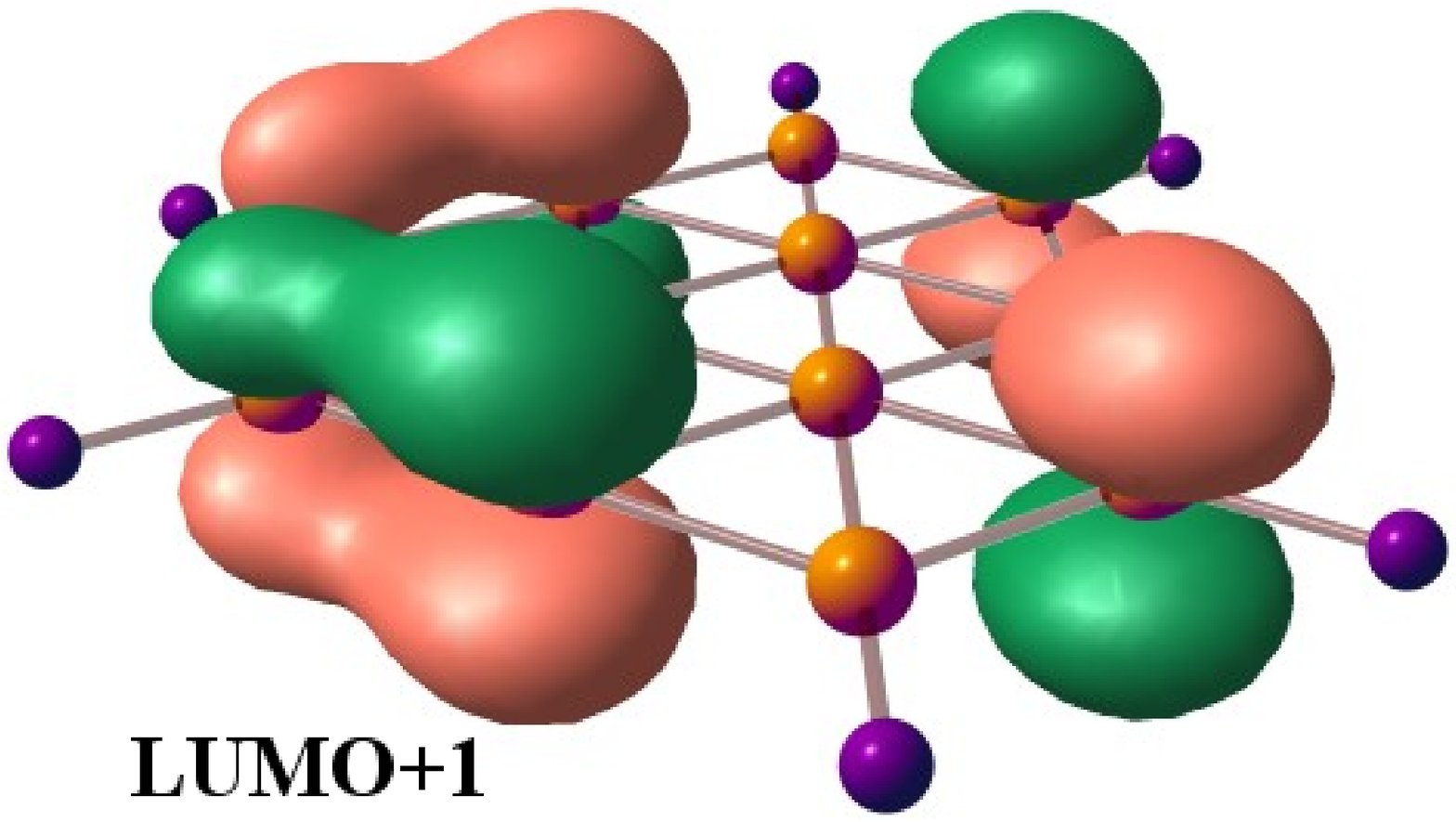}\label{fig:mos}

\caption{(Color online) Molecular orbitals (iso plots) of S-borozene from HOMO-1
to LUMO+1, obtained from the INDO-HF calculations.}

\end{figure}

Symmetry group $C_{2v}$ consists of four irreps labeled $a_{1}$,
$a_{2}$, $b_{1}$ and $b_{2}$. Assuming that the Cartesian $xy$-plane
lies in the plane of the molecule (\emph{cf}. Fig. \ref{Fig-struct}),
the symmetry adapted electronic structure of the HF ground state obtained
from our code is $1a_{1}^{2}2a_{1}^{2}1b_{1}^{2}3a_{1}^{2}2b_{1}^{2}4a_{1}^{2}1b_{2}^{2}3b_{1}^{2}5a_{1}^{2}4b_{1}^{2}6a_{1}^{2}7a_{1}^{2}8a_{1}^{2}5b_{1}^{2}2b_{2}^{2}1a_{2}^{2}9a_{1}^{2}$\\
$6b_{1}^{2}7b_{1}^{2}10a_{1}^{2}8b_{1}^{2}$, leading to $^{1}A_{1}$
many-particle configuration. Some of the lowest unoccupied orbitals
in the ascending order of energy are $3b_{2}$, $2a_{2}$, $4b_{2}$,
$11a_{1}$, $9b_{1}$\emph{ etc}. Of all the orbitals, the ones belonging
to $a_{2}$ and $b_{2}$ irreps are $\pi$ orbitals and $a_{1}$ and
$b_{1}$ are of $\sigma$ type. Thus we note that the highest occupied
molecular orbital (HOMO) is a $\sigma$ orbital, while the lowest
unoccupied molecular orbital (LUMO), is a $\pi$ orbital. For the
sake of visualization, some of the orbitals are given in Fig. \ref{fig:mos}.

While our LUMO orbital matches perfectly with that reported by Szwacki
\emph{et. al} \cite{szwacki}, however, we have a disagreement on
the nature of the HOMO orbital which was reported to be of $\pi$-type
by them\cite{szwacki}, but we obtain it to be of $\sigma$-type within
the INDO model. In order to rule out the possibility that this disagreement
could due to the use of the INDO model, or due to the S-borozene geometry,
we studied the nature of the HOMO obtained using the 6-311++g(d) basis
set both at RHF and DFT-B3LYP levels of \emph{ab initio} theory\cite{gaussian}
at various geometries. In all such DFT calculations, it turned out
to be a $\sigma$-type orbital, but\emph{ }for the RHF calculations,
the nature of HOMO was highly geometry sensitive with some geometries
yielding it to be $\sigma$-type, while others of $\pi$-type. This,
in our opinion, is a significant difference between the electronic
structures of benzene (whose HOMO and LUMO both are $\pi$ type),
and borozene. Thus, borozene, in our opinion is an example of both
$\sigma$ and $\pi$ conjugation because, close to the Fermi level,
occupied orbitals are mainly of $\sigma$ type, while the unoccupied
ones ($LUMO$, $LUMO+1$, $LUMO+2$) are of $\pi$ type.

Before presenting the results of our calculations of linear optical
absorption spectrum of S-borozene, we would like to discuss the influence
of geometry on the optical properties of borozene, considering the
fact that our calculations were performed on a highly symmetric conformer
S-borozene. For the purpose, we performed the optical absorption calculations
at the singles-CI (SCI) calculations both on the lowest energy structure
of borozene with unequal B-B bond lengths, and S-borozene. We found
that qualitatively, at the SCI level, spectra were very similar for
both the geometries, except that for S-borozene it was slightly red-shifted
compared to borozene. The first peak for borozene in this SCI calculation
was obtained at 2.6 eV in perfect agreement with excitation energy
of the first excited reported by Szwacki \emph{et al.}\cite{szwacki},
while that for S-borozene was located at 2.4 eV.

Next, we present and discuss the results of our INDO-MRSDCI based
optical absorption calculations on S-borozene. Comparisons are also
made with the absorption spectra of quasi-planar B$_{12}$ bare cluster,
and also with that of benzene. Because the ground state of the system
is $^{1}A_{1}$, as per electric-dipole selection rules, it can make
linear optical transitions to excited states belonging to symmetry
manifolds $^{1}A_{1}$, $^{1}B_{1}$, and $^{1}B_{2}$ through photons
polarized along the $y-$axis, $x-$axis, and $z-$axis, respectively.
Thus, transitions to $^{1}A_{1}$ and $^{1}B_{1}$ are through photons
polarized in the plane of the molecule ($xy$-plane in the present
case), while those to $^{1}B_{2}$ will be through perpendicularly
polarized photons. The INDO-MRSDCI calculations presented here involved
construction and diagonalization of very large CI matrices running
into millions of configurations, because simultaneously several states
were targeted. For example, for the $^{1}A_{1}$ symmetry manifold
the total number of configurations was more than 2.1 millions, while
those for $^{1}B_{1}$ and $^{1}B_{2}$ manifolds were in excess of
2.5, and 3 millions, respectively. Because of the large-scale nature
of these calculations, we believe that the results account for electron
correlation effects properly.

\begin{figure}
\includegraphics[width=8cm]{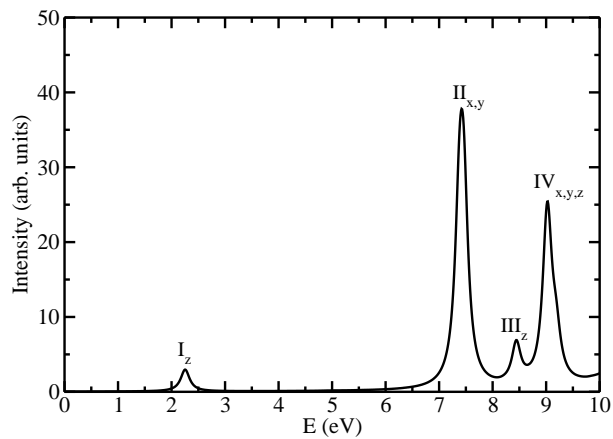}

\caption{Linear optical absorption spectrum of S-borozene, computed using the
INDO-MRSDCI approach. Important peaks, along with their polarization
characteristics, are labeled. A line width of 0.1 eV was used to compute
the spectrum.}
\label{Fig:spectrum-total}

\end{figure}

In Fig. \ref{Fig:spectrum-total} we present the combined optical
absorption spectrum of S-borozene, obtained from our INDO-MRSDCI calculations
are presented. As is obvious from the figures that the first absorption
feature is a small peak located at 2.25 eV, corresponding to a $z$-polarized
transition into a $B_{2}$ state. The many-particle wave function
of this peak consists mainly of two single excitations $H\rightarrow L+1$
and $H-1\rightarrow L$ ($H\equiv HOMO$; $L\equiv LUMO)$, both of
which are of $\sigma\rightarrow\pi^{*}$ type. After this feature
at 2.25 eV, the next peak follows after a large gap of $\approx5$
eV, and is due to two closely placed states at 7.38 eV ($B_{1}$-type)
and 7.55 eV ($A_{1}$-type). Thus, this intense peak can be reached
through photons which are polarized in the molecular plane, but not
through the $z$-polarized ones. The many-particle wave functions
of both these states are dominated by singly-excited configurations,
with the main configurations being $\vert H\text{\textrightarrow}L+3\rangle$
for the $B_{1}$ state, and $\vert H-1\text{\textrightarrow}L+3\rangle$
for the $A_{1}$ state. Thus, both these states correspond to $\sigma\rightarrow\sigma^{*}$
transitions.

Peak III of the spectrum, which is a weaker feature located at 8.44
eV, corresponds to a $B_{2}$ state, and can be reached through a
$z$-polarized photon. Many-particle wave function of this state is
a mixture of several singly-excited configurations which include $|H-4\rightarrow L\rangle$,
$|H-3\rightarrow L+1\rangle$, and $|H\rightarrow L+8\rangle$, all
of which are $\sigma\rightarrow\pi^{*}$ type excitations. Finally,
feature IV near 9.1 eV draws its oscillator strength from the transition
of the ground state to three closely spaced excited states of symmetries
$A_{1}$, $B_{1}$, and $B_{2}$ located at 9.03 eV, 9.01 eV, and
9.19 eV, respectively. The peak has mixed polarization features, out
of which $x$- and $y-$polarized transitions are more intense as
compared to the $z$-polarized one. The many-particle wave function
of the $A_{1}$ state is dominated by configurations $|H-1\rightarrow L+5\rangle$
and $|H\rightarrow L+4\rangle$, while that of $B_{1}$ state consists
mainly of $|H-1\rightarrow L+4\rangle$ and $|H\rightarrow L+5\rangle$,
all of which are of $\sigma\rightarrow\sigma^{*}$ type. The dominant
configurations in the wave function of the $B_{2}$ state are $|H\rightarrow L+8\rangle$
and $|H-1\rightarrow L+7\rangle$, which are of $\sigma\rightarrow\pi^{*}$
type. The absorption spectrum of S-borozene consists of several very
intense features beyond 10 eV as well, but we are not discussing them
here, given the fact that the ionization potential of this material
is estimated close to 9 eV\cite{szwacki}.

Because Szwacki \emph{et al.}\cite{szwacki} called borozene the boron
analogue of benzene, it is interesting to compare the optical absorption
spectra of the two materials. It is a well-known fact that the first
dipole allowed transition in benzene is from the $^{1}A_{1g}$ ground
state to the $^{1}E_{1u}$ excited state with the peak around 6.99
eV, through photons polarized in the plane of the molecule ($xy$-plane
in our case)\cite{benzene-exp}. This compares very well with peak
II (\emph{cf. }Fig. \ref{Fig:spectrum-total}) of S-borozene which
is also caused by photons polarized in the the $xy$-plane, and is
located around 7.5 eV. The only difference being that for the the
$^{1}E_{1u}$ state in benzene corresponds to a $\pi\rightarrow\pi^{*}$
transition, while the peak II of borozene corresponds to $\sigma\rightarrow\sigma^{*}$
transitions. This supports the hypothesis that borozene is both $\pi$
and $\sigma$ conjugated\cite{szwacki}. The difference between the
optics of the two materials is the presence of low energy feature
(peak I) in S-borozene corresponding to a $z$-polarized transition,
whose counterpart in benzene does not exist.

\begin{figure}
\includegraphics[width=8cm]{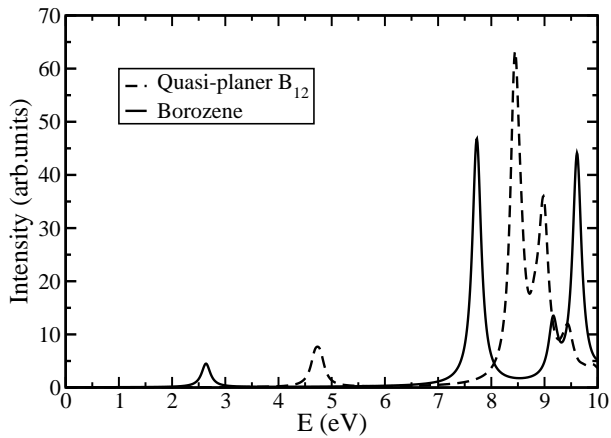}

\caption{ A comparison of the linear optical absorption spectra between S-borozene
(solid line) quasi-planer B$_{12}$ (broken line). The spectrum of
S-borozene is significantly red-shifted as compared to that of $B_{12}$.
A line width of 0.1 eV was used to compute the spectrum.}

\label{boroz_b12_compr}
\end{figure}

Recently, we studied the linear optical absorption in quasi-planer
$B_{12}$ cluster which has a convex shape with symmetry group $C_{3v}$\cite{sahu-b12},
and can be seen as a precursor of borozene\cite{szwacki}. Therefore,
a comparison of the optical properties of quasi-planar $B_{12}$ and
borozene can help us understand the influence of hydrogen passivation
on the optical properties of $B_{12}$. In Fig. \ref{boroz_b12_compr},
the linear optical absorption spectra of the two clusters is presented,
and it is obvious that the first few peaks of S-borozene occur at
lower energies as compared to the quasi-planar $B_{12}$. Moreover,
the $H\rightarrow L$ transition is symmetry forbidden in S-borozene,
which is not the case for quasi-planar $B_{12}$, and indeed contributes
to the wave function of the first peak in its spectrum. Thus the main
contribution of hydrogen passivation on the optical properties is
that it makes the structure completely planar (as against quasi-planar)
leading to conjugation and higher symmetry. Lower energy gap of S-borozene
is most certainly a consequence of conjugated nature of electrons
in the system. 

In order to explore aromaticity of borozene, Szwacki \emph{et al.}\cite{szwacki}\emph{
}computed, and discussed its NICS plots, and anisotropy of magnetic
susceptibility (AMS). With a similar aim we computed the three diagonal
Cartesian components of the static polarizability tensor, $\alpha_{xx}$,
$\alpha_{yy}$, and $\alpha_{zz}$ of borozene both at the INDO-HF
as well as INDO-SDCI levels. For the purpose we adopted a finite-field
approach, and computed the second derivative of the total energy of
the system numerically, with respect to the three Cartesian components
of the external electric field, whose values was taken to be $0.005$
atomic units. The numerical values obtained in our calculations correspond
to valence electron contribution to the polarizability because of
the nature of the INDO Hamiltonian. However, our aim here is not to
obtain exact numerical values of the polarizability tensor, rather
to compare the values of its various components. It is obvious from
the results of our calculations presented in table \ref{tab:polariz}
that: (a) inclusion of correlation effects does not change the values
of $\alpha_{ii}$ very much, and (b) the values of in-plane components
($\alpha_{xx}$ and $\alpha_{yy}$) are significantly larger than
the perpendicular component $\alpha_{zz}$. \emph{Ab initio} calculations
performed on conjugated molecules like benzene,\cite{lazzeretti-alpha,hinch-alpha}
anthracene\cite{perez-alpha}, and ethane,\cite{hinch-alpha} exhibit
similar anisotropy in the components of the polarizability tensor
. And, indeed, based upon the magnitude of this {}``in-plane'' vs.
{}``perpendicular'' anisotropy, Lazzeretti and Tossell\cite{lazzeretti-alpha}
argued for an {}``aromaticity scale'' to characterize $\pi$-conjugated
systems. Using the same logic, the same anisotropy found in our calculations
clearly implies the aromaticity in borozene, but as a consequence
of both $\pi$ and $\sigma$ conjugation in the plane of the molecule.
This fact is also obvious from the optical absorption spectrum of
the material in which intensities of the $z$-polarized peaks are
much smaller than those of $x$- and $y$-polarized ones.

\begin{table}
\caption{Values of the components of static dipole polarizability tensor of
borozene computed using the INDO-HF and INDO-SDCI methods (in atomic
units).}

\begin{tabular}{|c|c|c|}
\hline 
Method & $\alpha_{xx}=\alpha_{yy}$ & $\alpha_{zz}$\tabularnewline
\hline
\hline 
INDO-HF & 34.273 & 21.833\tabularnewline
\hline 
INDO-SDCI & 33.705 & 22.119\tabularnewline
\hline
\end{tabular}

\label{tab:polariz}
\end{table}

\section{Conclusions}

\label{sec:conclusions}

In conclusion, a correlated study of electronic structure and optical
properties of a recently proposed boron-hydrogen cluster, borozene,
was presented using the semi-empirical MRSDCI-INDO method. Our calculations
of the linear optical absorption spectrum can be used to characterize
this substance in future optical absorption experiments. We also calculated
the diagonal components of the static polarizability tensor $\alpha_{ij}$
of the cluster, and found significant larger values for the components
along the plane of the molecule, as compared to the one perpendicular
to it. This anisotropy in the components of the polarizability tensor
implies that the electrons are quite delocalized along the plane of
the molecule leading to a considerably larger response to an electric
field directed along the plane, as compared to that perpendicular
to the plane. Therefore, we argue that this anisotropy in the static
dielectric polarizability of the cluster is an evidence of its aromaticity,\cite{lazzeretti-alpha}
caused both by $\sigma$ and $\pi$ conjugation along the plane of
the molecule.
\begin{acknowledgments}
We gratefully acknowledge useful private communications with Dr. C.
J. Tymczak in which he clarified the optimized geometry obtained in
their calculations on borozene\cite{szwacki}.\end{acknowledgments}

\end{document}